# Quasiparticle interference on cubic perovskite oxide surfaces

Yoshinori Okada[1,*], Tay-Rong Chang[2], Guoqing Chang[3,4], Ryota Shimizu[1],

Shiue-Yuan Shiau[3,4], Horng-Tay Jeng[2,5], Susumu Shiraki[1], Arun Bansil[6], Hsin Lin[3,4], Taro Hitosugi[1,7]

[1]Advanced Institute for Materials Research (AIMR), Tohoku University, Sendai 980-8577, Japan

[2]Department of Physics, National Tsing Hua University, Hsinchu 30013, Taiwan

[3]Centre for Advanced 2D Materials and Graphene Research Centre,

National University of Singapore, Singapore 117546

[4] Department of Physics, National University of Singapore, Singapore 117542

[5] Institute of Physics, Academia Sinica, Taipei 11529, Taiwan

[6] Department of Physics, Northeastern University, Boston, Massachusetts 02115, USA

[7] Department of Applied Chemistry, Tokyo Institute of Technology, Tokyo 152-8552, Japan

*e-mail: yoshinori.okada@wpi-aimr.tohoku.ac.jp

We report the first observation of coherent surface states on cubic perovskite oxide SrVO$_3$(001) thin films through spectroscopic imaging scanning tunneling microscopy. A direct link between the observed atomic-scale interference patterns and the formation of a d$_{xy}$-derived surface state is supported by first-principles calculations. Furthermore, we show that the apical oxygens on the topmost VO$_2$ plane play a critical role in controlling the spectral weight of the observed coherent surface state.

Establishing a microscopic understanding of the mechanism leading to coherent two-dimensional (2D) states at surfaces/interfaces of transition metal perovskite oxides is a key step for further exploration of exotic quantum states in various systems [1-5]. Despite extensive investigations of the electronic surface states in transition metal perovskite oxides using angle-resolved photoemission spectroscopy (ARPES) [6-10], the underlying microscopic mechanism that leads to the formation of coherent electronic 2D states has remained elusive. In this connection, the link between the surface 2D states and their atomic origins must be unraveled in order to make progress. Scanning tunneling microscopy (STM) and scanning tunneling spectroscopy (STS) allow us to observe the surface interference pattern at the atomic scale, and provide a probe of 2D quasiparticle states with well-defined wave vectors [11-14]. However, perovskite materials are intrinsically three-dimensional (3D) crystals, and it is difficult to obtain atomically well-aligned surfaces by conventional cleaving techniques.

In this Letter, we report the first STM/STS observation of quasiparticle interference (QPI) patterns on the epitaxial film surfaces of perovskite oxide $SrVO_3$ (001). The $SrVO_3$ bulk has simple cubic symmetry with one electron in the 3d state, and hence it has been used as a prototype system for understanding correlated transition metal perovskite oxides [15-29]. We interpret our experimental STM/STS results through parallel first-principles calculations to establish a microscopic understanding of the appearance and disappearance of observed coherent surface 2D quasiparticle states.

The epitaxial $SrVO_3$(001) thin films were grown on a buffered-HF-etched Nb-doped (0.05 wt.%) $SrTiO_3$(001) substrate by using pulsed laser deposition (PLD). In the three samples (thicknesses: 20, 40, and 110 nm) prepared for this study [30], the X-ray diffraction patterns showed negligible epitaxial strain. We used an STM equipped with PLD to transfer the as-deposited thin films to a low-temperature (4 K) STM head under ultra-high vacuum conditions [31]. Figure 1(a) presents an STM topographic image of the $SrVO_3$(001) surface showing a square lattice of protrusions with randomly distributed defects. The periodicity of the protrusions, shown in Fig. 1(b), is approximately 5.5 Å and exhibits a $(\sqrt{2}\times\sqrt{2})$-$R45°$ surface reconstruction everywhere on the scanned surfaces.

The √2×√2 superlattice surface structure seen in Fig. 1(c) is formed by a VO$_2$-terminated layer with apical oxygen adsorption [32,33]. Naively, we would expect the topmost V atoms to have the same valence as the bulk V$^{4+}$ atoms because the bulk structure can be seen as stacked (Sr$^{2+}$O$^{2-}$)$^0$ and (V$^{4+}$O$^{2-}_2$)$^0$ layers. However, previous emission-angle-dependent photoemission studies show more V$^{5+}$ atoms on the surface [22, 34], which means that the V atoms near the surface are more oxidized than those in the bulk. This observation excludes the possibility of a *bare* VO$_2$ termination with √2×√2 buckling near the surface since such a surface structure is unlikely to accommodate substantially oxidized V atoms. A 50% coverage of the apical oxygen sites (forming a SrVO$_{3.5}$ surface structure) is however allowed, which in turn would lead to the formation of the √2×√2 superlattice. Similar √2×√2 superstructures with 50% coverage of the apical oxygen sites have also been seen in manganite perovskites [35-38].

Figure 1(d) shows spatially-averaged differential conductance (d$I$/d$V$) spectra. The metallic electronic properties observed on the surface are consistent with previous ARPES experiments [20]. The d$I$/d$V$ spectra further show a large spectral weight at zero sample bias voltage $V_s$, a peak at $V_s$ = +400 mV, and two minima at $V_s$ = −410 mV and +710 mV. Right at the peak position ($V_s$ = +400 mV), a shoulder structure has also been observed in angle-integrated inverse photoemission experiments [39]. The energy at −410 mV for the conductance minimum agrees quantitatively with the band-edge position at the $\bar{\Gamma}$ point in previous ARPES studies [40].

Figures 2(a) and 2(b) show the topographic and conductance (d$I$/d$V$) images which were simultaneously measured at $V_s$= −150 mV. In order to extract the periodic QPI pattern, which is present only in the conductance image, we compared the Fourier transforms (FTs) of the two images, which are shown in Figs. 2(c) and 2(d). While both FT images show peaks at √2×√2 spots [circles in Fig. 2(c)], only the conductance FT image exhibits an additional periodicity with respect to the momentum $q^*$s [dashed ellipses in Fig. 2(d)]. Figure 2(e) shows the energy-dependent intensity profile of the FT images along the $\bar{\Gamma}\bar{X}$ direction for the three samples. The clear energy dispersion seen in Fig. 2(e) provides indisputable evidence for the existence of coherent electronic states on the SrVO$_3$(001) surface.

In order to understand the energy dependence of the QPI pattern, we constructed a semi-infinite slab model using Wannier functions derived from first-principles density functional theory (DFT) based calculations; see Supplementary Materials for computational details [30]. We refer to an O-adsorbed V-atom site as "$V_O$-site" and to a bare V-atom site without apical oxygen as "$V_b$-site". Figure 3(b) shows the calculated orbital-dependent densities of states (DOS) as well as an average DOS between $V_O$- and $V_b$- sites. Using a bandwidth renormalization factor of 0.4, which is close to the value obtained from previous ARPES experiments [20] and first principles calculations [21], the computed total DOS on the surface [Fig. 3(b)] reproduces the conductance peak (+400 mV) and the band-edge position (−410 mV) in the experimental d$I$/d$V$ spectrum [Fig. 1(d)]. We will hereafter denote the renormalized energy bands by $E_{re}$. Such a renormalization of the DFT bands, despite strong correlation effects are underestimated, provide a reasonable approach for capturing the orbital information needed to model the emergent QPI patterns.

In order to establish that the QPI pattern is directly linked to the formation of the $d_{xy}$-derived coherent 2D band on the surface, we project the spectral weight of $d_{xy}$ and $d_{xz}$/$d_{yz}$ derived bands at $V_O$- and $V_b$-sites onto the surface Brillouin Zone (BZ) [Fig. 3(c)]. Results along the high-symmetry lines are shown in Figs. 3(d)-(g). For both $V_O$- and $V_b$- sites, the $d_{xy}$-derived spectral weight shows a clear energy dispersion [Figs. 3(d)-3(f)], indicating that the $d_{xy}$-derived bands feature a strong 2D nature on the (001) surface. At lower energies, this $d_{xy}$-derived 2D spectral weight becomes the dominant electronic component at both $V_O$- and $V_b$- sites [see shaded region on the top horizontal axis in Fig. 3(b)], where a clear QPI pattern is observed. Since interlayer coupling between in-plane extended $d_{xy}$ orbital is small and its surface onsite energy is higher from bulk [0.1eV and 0.3 eV higher for $V_O$- and $V_b$- sites, respectively, see Fig. 3(k)], we conclude that the emergent QPI pattern on the surface originates from $d_{xy}$-derived quasiparticle *surface states*, which are more or less isolated not only from the *surface* $d_{xz}$/$d_{yz}$ states but also from the *bulk* states.

Our calculations also capture the appearance of the momentum $q^*$ selectively along the $\overline{\Gamma X}$ direction. We compute the scattering probability $I(q)$ in terms of the spectral weight $A(k)$ using $I(q) = \oint A(k) \cdot A(k+q) \mathrm{d}k$. Typical results for $A(k)$ and $I(q)$ by considering $E_{re} = -250$ meV are shown in Figs. 3(h) and 3(i), respectively. Because the shape of $A(k)$ deviates from a simple circle, an increase in the joint-DOS occurs along the $\overline{\Gamma X}$ direction in $I(q)$. The magnitude and direction of this enhanced joint-DOS is consistent with the value of $q^*$ obtained from the experimental FT of d$I$/d$V$ image at the corresponding energy $eV_s$= -250meV [Fig. 3(j)].

As the energy increases across the $E_F$, the QPI signal gradually decreases in intensity and disappears altogether ultimately [Fig. 2(e)]. This fading of the QPI signal above $E_F$ contradicts the behavior of simple metallic surfaces in which, according to the conventional Fermi liquid theory, quasiparticle states should have a longer lifetime near $E_F$, and as a result, the QPI patterns should be more prominent [41]. We consider three possible scenarios in this connection as follows: (1) Suppression (decoherence) of the quasiparticle interference via inelastic electron-electron, electron-phonon and electron-plasmon scatterings. However, the symmetric energy dependence on both sides of $E_F$ expected for these inelastic scattering processes contradicts the monotonic fading of the QPI signal observed across $E_F$; (2) Effect of the shape of $A(k)$. This scenario however is also unlikely because the shape of the calculated $A(k)$ changes from circle-like (smaller joint-DOS) to square-like (larger joint-DOS) across the $E_F$, which completely fails to explain the experimentally observed fading of the QPI signal; and (3) Suppression of the $d_{xy}$-derived spectral weight itself [42,43]. Indeed, our calculations show a substantial suppression of the $d_{xy}$-derived spectral weight, especially at the $V_b$-site [dotted squares in Fig. 3(f)]. This is driven by two mechanisms: (i) inter-orbital spectral weight transfer and (ii) intra-orbital spectral weight transfer due to sub-lattice formation.

The inter-orbital spectral weight transfer mechanism, which involves transfer of spectral weight from the $d_{xy}$ to $d_{xz}/d_{yz}$ orbitals at the $V_b$-site, is illustrated schematically in Fig. 4(a). As the axially-extended $d_{xz}/d_{yz}$-derived *bulk* bands have an intrinsically strong dispersion normal to the (001) surface, the projection of such a band onto the (001) surface BZ results in a broad distribution of the spectral weight in momentum space. At the $V_b$-site, the broad spectral weight distribution shown in Fig. 3(g) indicates the existence of a strong coupling between the $d_{xz}/d_{yz}$-derived *surface* and *bulk* bands. Such a strong mixing can be naturally understood from the smaller bulk-surface energy separation (~0.2 eV) for $d_{xz}/d_{yz}$ orbital at the $V_b$-site compared to that for the $V_o$-site (~1.2 eV) [see Fig. 3(k)]. An important consequence is the possible coexistence of the $d_{xy}$-derived and $d_{xz}/d_{yz}$-derived bands at the same energy $E$ and momentum $k$ over a wide range [dotted squares in Fig. 3(f)].

This eventually leads to the suppression of the $d_{xy}$-derived spectral weight via hybridization with the $d_{xz}/d_{yz}$-derived bands.

We next discuss the intra-orbital spectral weight transfer mechanism, which involves transfer of spectral weight from the $d_{xy}$ orbital at the $V_b$-site to the $d_{xy}$ orbital at the $V_O$-site as shown schematically in Fig. 4(b). The calculated on-site energies for the $d_{xy}$ orbital at the $V_O$- and $V_b$- sites are ~0.1 and ~0.3 eV, respectively. In general, a difference in on-site energies will induce a spectral weight transfer between these inequivalent sites around the zone boundary [44]. We confirm this effect by focusing on the lower energy band branch near the $\bar{X}$ point [dotted squares in Figs. 3(d) and 3(f)] since this is the main electronic component that creates the QPI signal. Our calculations show that the spectral weight of the $d_{xy}$ state at the $V_b$-site (with higher potential) for the lower band branch is suppressed, while the spectral weight of the $d_{xy}$ state at the $V_O$-site (with lower potential) for the same $(E, k)$ region around the $\bar{X}$ point is simultaneously enhanced.

The discussion above of the inter-orbital and intra-orbital spectral weight transfer shows that it is the $d_{xy}$-derived spectral weight distribution that causes the QPI signal on the $(\sqrt{2}\times\sqrt{2})$-$R45^o$ surface to fade monotonically with energy [see Fig. 4(c)]. At low energies, we obtain a uniform distribution from the $d_{xy}$-derived spectral weight [see Fig. 4(d)], while at higher energies, the suppressed spectral weight from the $d_{xy}$-band at the $V_b$-site [Figs. 4(a) and 4(b)] disrupts the uniformity of the $d_{xy}$-derived spectral weight distribution [see Fig. 4(c)].

Our results yield important insights for designing 2D quasiparticle states at surfaces/interfaces of perovskite oxides. Most importantly, we identify the pivotal role of apical oxygens in driving 2D electronic states. Because of the large on-site energy difference at the $V_O$-site, the negatively-charged apical oxygen strongly attracts holes and eliminates electrons from the axially-extended $d_{xz}$ and $d_{yz}$ states [see Fig. 3(k) (left)]. The apical oxygen thus plays a key role in isolating the in-plane $d_{xy}$ and out-of-the-plane $d_{xz}/d_{yz}$ orbitals on the (001) surface. Since apical oxygens are ubiquitous in perovskite oxides, these considerations can also be applicable for designing $e_g$ systems such as the cuprates, where the interplay between the $d_{x2-y2}$ and out-of-the-plane orbitals including $d_{z2}$ would be important [45]. On the other hand, an obvious unique factor in SrVO$_3$(001) is its electron filling of one electron in the *bulk* 3d system. Interestingly, our results suggest that the SrVO$_3$(001) surface is similar to the 2D metallic state at SrTiO$_3$-based surfaces/interfaces because in both systems the degeneracy of the t$_{2g}$ orbitals is lifted and the d$_{xy}$-derived band becomes dominant at lower energies [46-58]. This similarity suggests that the SrVO$_3$(001) surface is a promising candidate for exploring emergent quantum phases involving magnetism and superconductivity.

**Acknowledgements**

We thank H. Kumigashira, M. Kobayashi, H. Ishida, and K. Sato for helpful discussions, and D. Walkup and P. Han for critical reading of the manuscript. Y. O. acknowledges funding through the Japan Society for the Promotion of Science grant-in-aid numbers 26707016 and 25886004. T. H. acknowledges funding through the Japan Society for the Promotion of Science grant-in-aid numbers 26246022, 26106502, JST-CREST, and JST-PRESTO. This work was supported by the World Premier Research Center Initiative, promoted by the Ministry of Education, Culture, Sports, Science and Technology of Japan. The work at Northeastern University was supported by the US Department of Energy (DOE), Office of Science, Basic Energy Sciences grant number DE-FG02-07ER46352 (core research), and benefited from Northeastern University's Advanced Scientific Computation Center (ASCC), the NERSC supercomputing center through DOE grant number DE-AC02-05CH11231, and support (applications to layered materials) from the DOE EFRC: Center for the Computational Design of Functional Layered Materials (CCDM) under DE-SC0012575. H. L. acknowledges the Singapore National Research Foundation for support under NRF award number NRF-NRFF2013-03.

**Author contributions**

Y. O. pursued experiments including film growth, STM/STS measurements and data analysis. T. R., C, G. C., H. T. J., S. S., A. B., and H. L. performed theoretical calculations. Y. O., R. S., S. S., and T. H. discussed experimental results. All authors discussed the conclusions of this paper. Y. O., T. H., H. L., and A. B. wrote the paper.

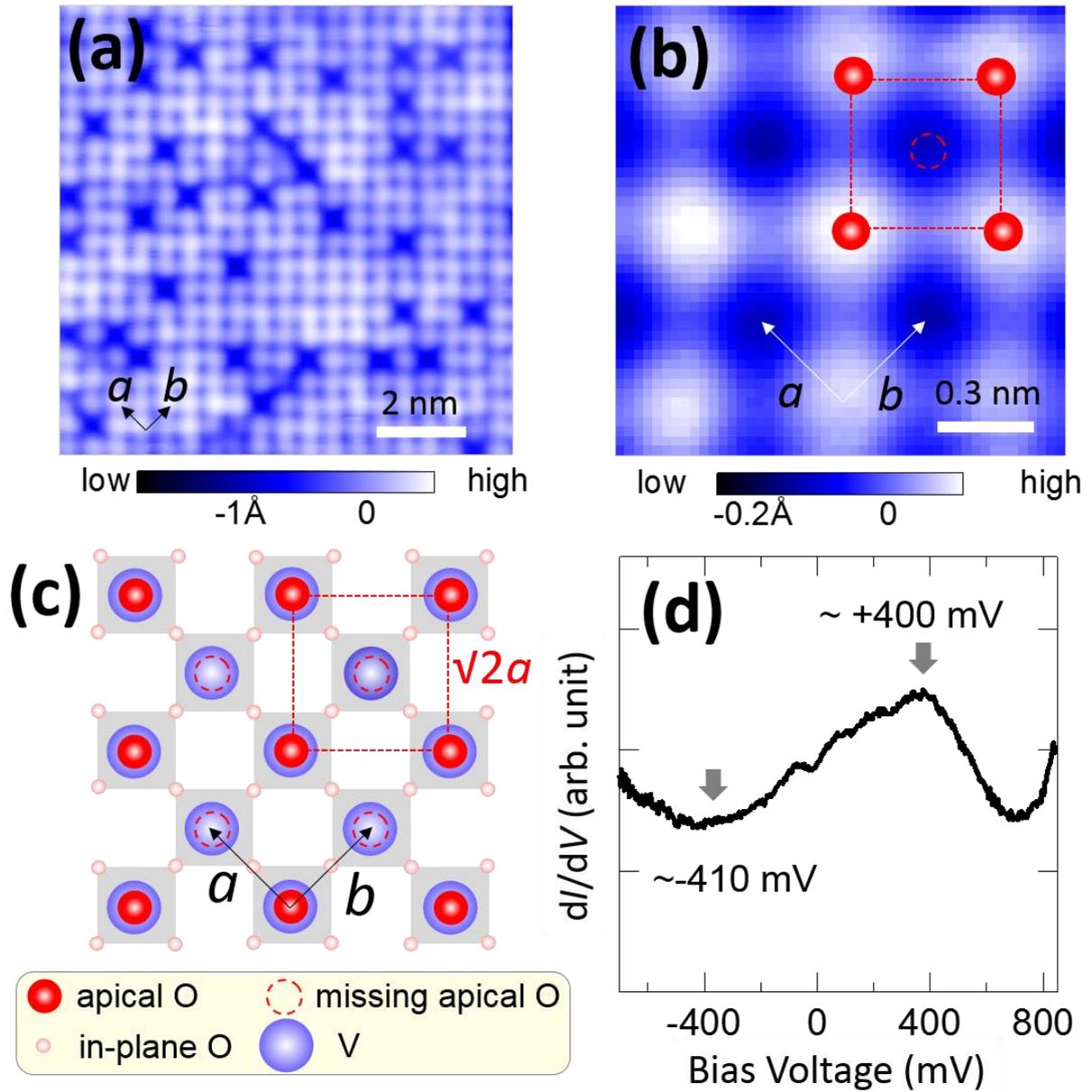

**FIG. 1 (color online)** Typical topographic images of SrVO$_3$(001) surface and metallic tunneling spectra (110-nm-thick film). **(a), (b)** Topographic images with sample bias voltage $V_s = -100$ mV **(a)**; the enlarged image in **(b)** shows the ($\sqrt{2}\times\sqrt{2}$)-$R45°$ structure. **(c)** The model used to generate the topographic image with a ($\sqrt{2}\times\sqrt{2}$)-$R45°$ structure. **(d)** Spatially-averaged tunneling spectra (d$I$/d$V$) obtained from the surface.

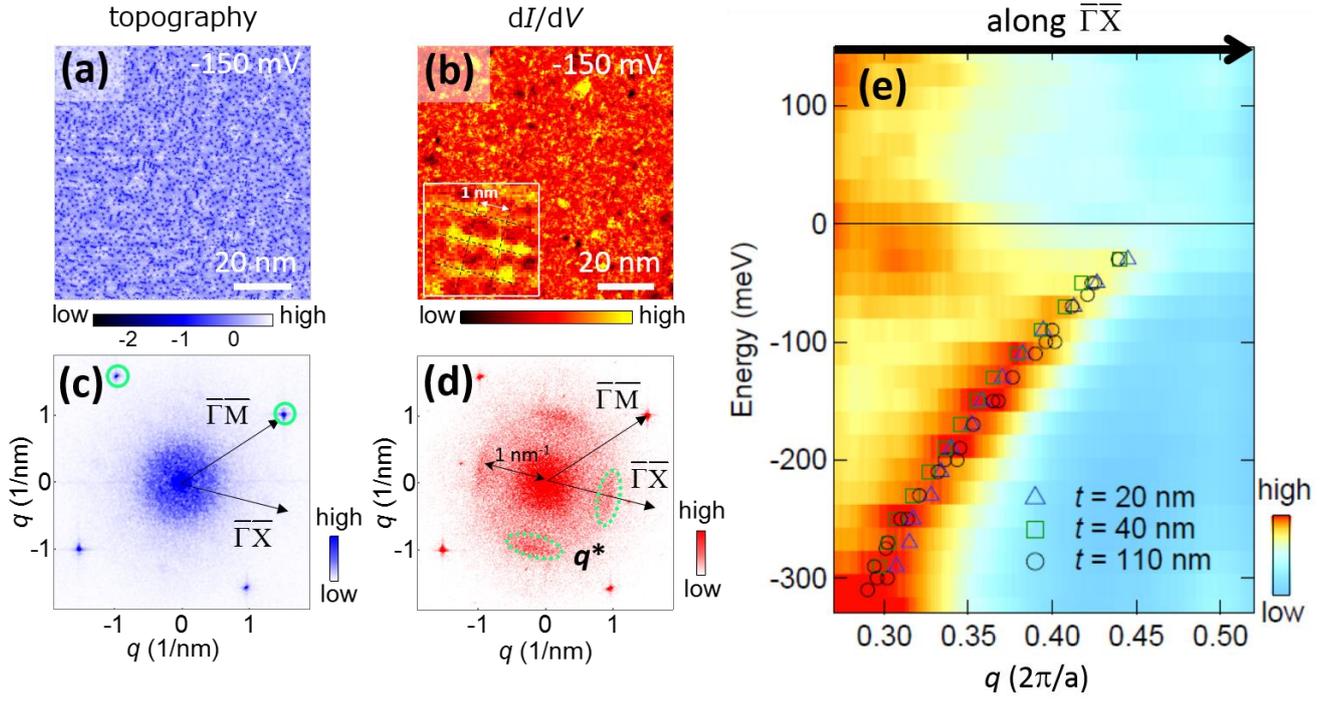

**FIG. 2 (color online)** Observation of quasiparticle interference (QPI) on the surface of SrVO$_3$(001). **(a)**, **(b)** Simultaneously acquired topographic **(a)** and conductance **(b)** images at $V_s = -150$ mV. The inset in **(b)** shows the enlarged image in which we can see a wave-like pattern with approximately 1 nm periodicity. **(c)**, **(d)** FTs of **(a)** and **(b)**, respectively. Bragg peaks representing the $\sqrt{2}a \times \sqrt{2}a$ apical oxygen structure are highlighted by green circles in (c), and the wave-vector $\boldsymbol{q^*}$ is highlighted by dashed-green ellipses in (d). $\overline{\Gamma X}$ direction corresponds to the direction of the nearest-neighbor V atoms. The data shown in **(a)**-**(d)** are taken from a 110-nm-thick film. **(e)** Energy-dependent intensity profiles of the FTs along $\overline{\Gamma X}$ (averaged intensity over three samples with thickness 20 nm, 40nm, and 110 nm), along with the values of $q^*$ obtained from the three films. The QPI patterns were independent of film thickness (Fig. S2), which is consistent with fully relaxed films.

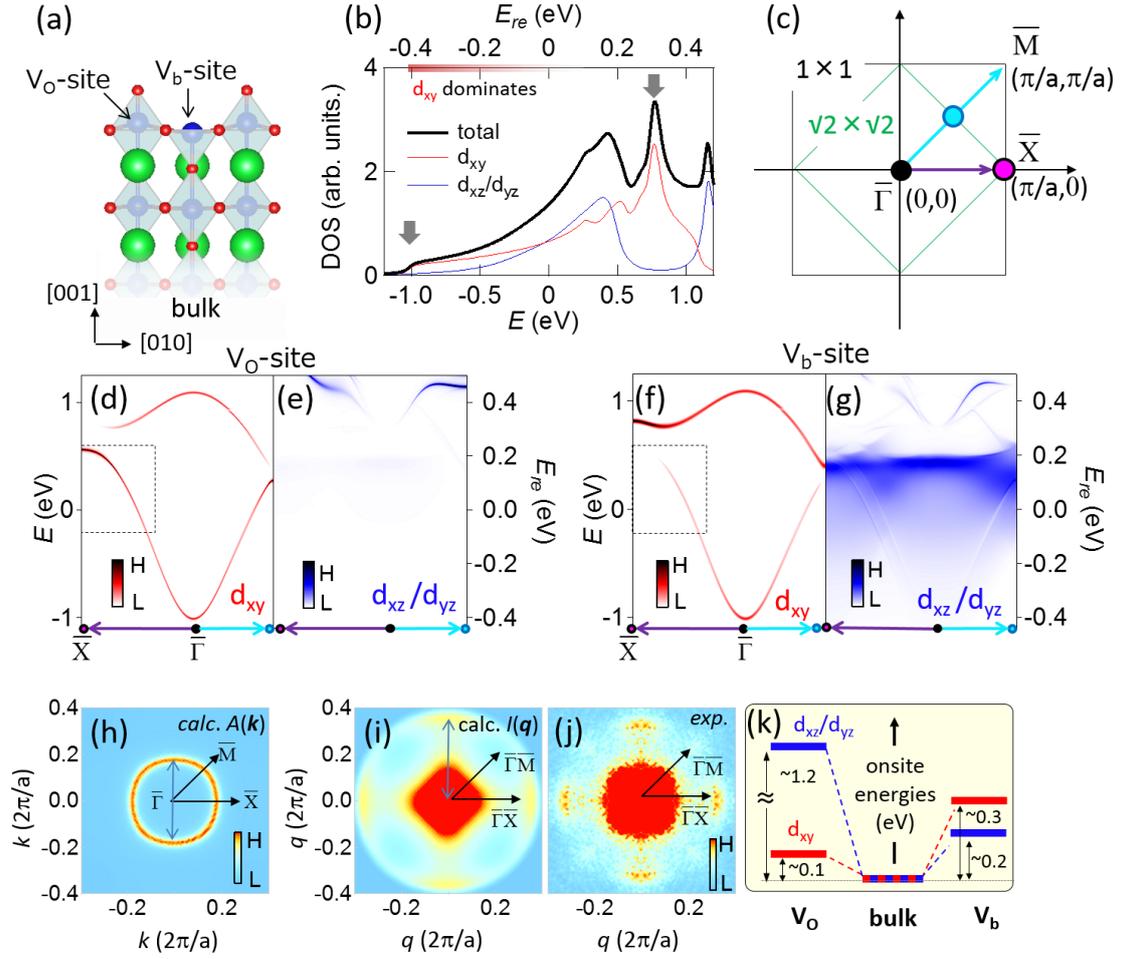

**FIG. 3 (color online)** Simulation of the electronic states on the surface of SrVO$_3$(001). **(a)** Cross-sectional view of the calculated relaxed near surface structure. V$_O$ and V$_b$ denote the V-atom sites with and without apical oxygen, respectively. **(b)** Calculated density of states averaged over V$_O$- and V$_b$-sites. **(c)** Schematic drawing of the surface Brillouin Zone (BZ). Green lines correspond to the BZ of ($\sqrt{2}\times\sqrt{2}$)-$R45°$ structure. **(d)**–**(g)** Spectral weights of the d$_{xy}$ and d$_{xz}$/d$_{yz}$ orbitals for V$_O$ (V$_b$) are shown in **(d)** and **(e)** (**(f)** and **(g)**), respectively. The top axis of (**b**) and right axes of (**d**)–(**g**) represent renormalized energy scale $E_{re}$ using a renormalization factor of 0.4 (*i.e.*, $E_{re} = 0.4E$). **(h)**-**(j)** Simulated total spectral weight at $E_{re} = -250$ mV (**h**) and the associated scattering probability (**i**). The experimental Fourier transform image of conductance mapping at $V_s=-250$ mV is shown in (**j**) for comparison with (**i**). **(k)** The orbital- and site-dependence of onsite energies relative to the degenerate t$_{2g}$ states in bulk states; see Table S2 for details. The onsite energy differences within our model mainly reflect local crystal field effects; these are not related to onsite Coulomb repulsion $U$. See [30] for details of our calculations.

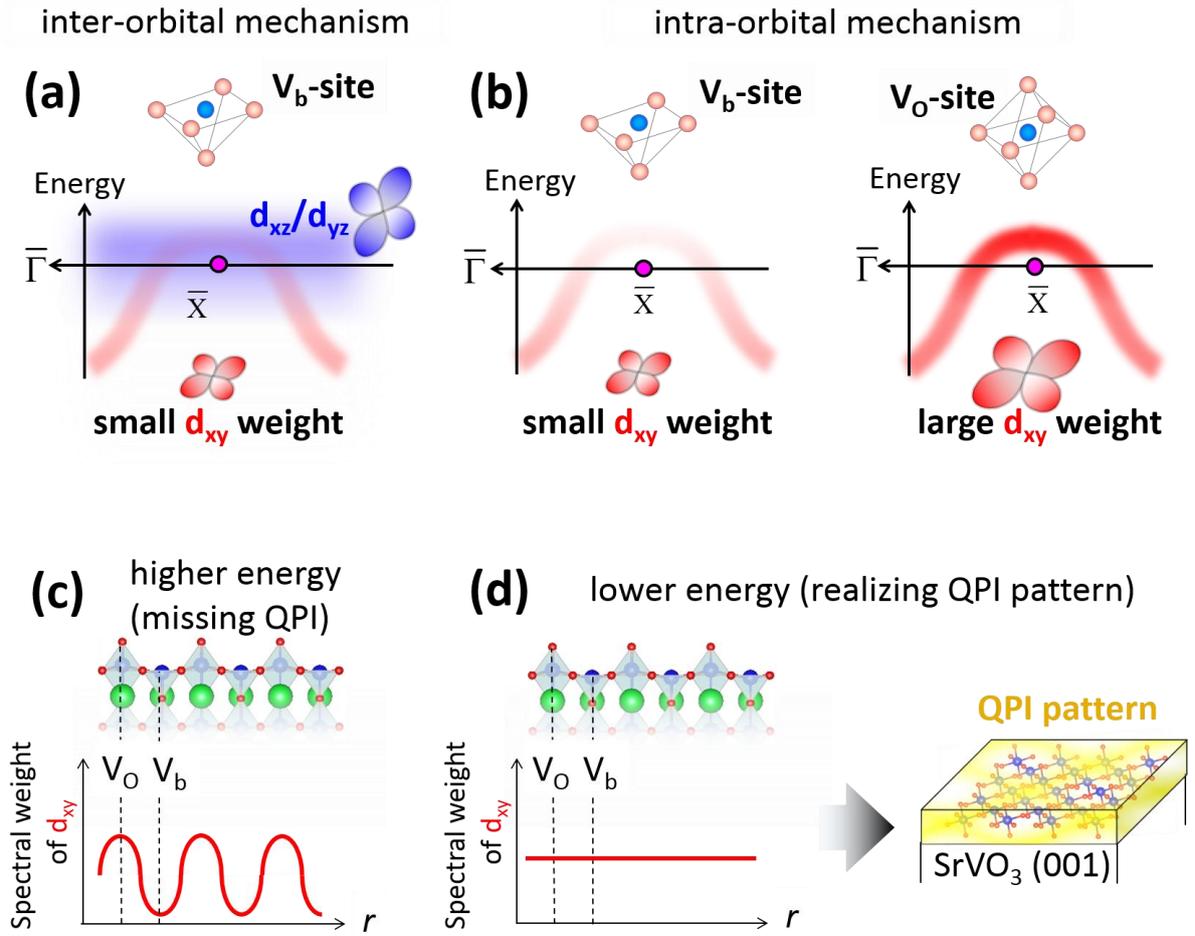

**FIG. 4 (color online)** Schematic illustration of inter-orbital and intra-orbital spectral weight transfer mechanisms and the related appearance/disappearance of QPI patterns. **(a)** Spectral weight transfer from $d_{xy}$ orbitals (red) to $d_{xz}/d_{yz}$ orbitals (blue) at the $V_b$-site. **(b)** Spectral weight transfer from $d_{xy}$ orbitals (red) at the $V_b$-site (left) to $d_{xy}$ orbitals (red) at the $V_O$-site (right). **(a)** and **(b)** only highlight the lower energy band branch near the $\bar{X}$ point. **(c)**, **(d)** Distribution of $d_{xy}$-derived spectral weight (red curve) in real space for the high- **(c)** and low- **(d)** energy regions.

# Quasiparticle interference on cubic perovskite oxide surfaces

Yoshinori Okada[1,*], Tay-Rong Chang[2], Guoqing Chang[3,4], Ryota Shimizu[1],

Shiue-Yuan Shiau[3,4], Horng-Tay Jeng[2,5], Susumu Shiraki[1], Arun Bansil[6], Hsin Lin[3,4], Taro Hitosugi[1,7]

[1] Advanced Institute for Materials Research (AIMR), Tohoku University, Sendai 980-8577, Japan

[2] Department of Physics, National Tsing Hua University, Hsinchu 30013, Taiwan

[3] Centre for Advanced 2D Materials and Graphene Research Centre,

National University of Singapore, Singapore 117546

[4] Department of Physics, National University of Singapore, Singapore 117542

[5] Institute of Physics, Academia Sinica, Taipei 11529, Taiwan

[6] Department of Physics, Northeastern University, Boston, Massachusetts 02115, USA

[7] Department of Applied Chemistry, Tokyo Institute of Technology, Tokyo 152-8552, Japan

e-mail:* yoshinori.okada@wpi-aimr.tohoku.ac.jp

**Supplementary Figures**

**Figure S1 |** Thin film growth of $SrVO_3$(001) on Nb-doped $SrTiO_3$(001).

**Figure S2 |** *Ex situ* XRD characterization.

**Figure S3 |** Typical STM images of $SrVO_3$(001) film surfaces.

**Figure S4 |** Energy evolution of conductance maps and their FTs (110–nm-thick film).

**Figure S5 |** Energy evolution of the quasiparticle interference (QPI) patterns for samples with thicknesses of 20 nm, 40 nm, and 110 nm.

**Supplementary Tables**

**Table S1 |** Experimental sample properties.

**Table S2 |** On-site energies for the five orbitals studied for the surface and bulk.

## Methods

**Epitaxial thin-film growth and *ex situ* characterization**

SrVO$_3$(001) films were grown on buffered-HF-etched Nb-doped (0.05 wt %) SrTiO$_3$(001) substrates supplied by Shinkosha Corp. Lattice constants for the substrate and SrVO$_3$ are 0.3905 nm and 0.384 nm[1-3], respectively, and the lattice mismatch is approximately 1.6%. The base pressure of the thin-film growth chamber was approximately 5×10$^{-10}$ Torr. Pulsed-laser deposition (PLD) processes employing a KrF laser ($\lambda$ = 248 nm) with a repetition rate of 2 Hz and a fluence of 1 J/cm$^2$ were used for the ablation of the target. A nearly single-phase polycrystalline Sr$_2$V$_2$O$_7$ pellet (Kojundo Chemical Laboratory Corp.) was used as the target. During the film deposition and annealing processes, substrates were resistively heated and temperature was monitored using a pyrometer. Before film deposition, the substrate was degassed at 500 °C in oxygen with a partial pressure $P_{O2}$ < 1×10$^{-8}$ Torr (Fig. S1). Following degassing, the films were deposited on the substrates while monitoring the film surface using reflection high-energy electron diffraction (RHEED) (Fig. S1). The films were deposited at 750 °C; this temperature is similar to temperatures used in previous studies of SrVO$_3$ thin-film growth, which ranged from 600–850 °C[4-7]. Identical growth conditions (Fig. S1), except for the film deposition time, were used for all three films investigated (Fig. S2 and Table S1).

**Film thickness measurements**

Film thicknesses were estimated by dividing the deposition time by the periodicity of the RHEED intensity oscillations observed at the initial stage of film deposition. Film thicknesses were also measured *ex situ* using a DEKTAK 3030ST mechanical profiler (veeco.com). The thickness values obtained from the two methods were consistent within experimental errors.

***Ex situ* X-ray diffraction characterizations**

All investigated films were characterized by *ex situ* X-ray diffraction (XRD) after taking scanning tunneling microscopy (STM) measurements. The *c*-axis lattice constant (Figs. S2a-d) and the rocking-curve full-width-at-half-maximum (FWHM) of 004 peaks (Fig. S2) were determined using out-of-the-plane XRD measurements.

**Spectroscopic imaging using scanning tunneling microscopy**

A low temperature ultra-high vacuum scanning tunneling microscope equipped with a PLD system[8] was employed. In this way, the thin films could be transferred without exposing their surfaces to air. STM measurements were performed at 4 K using electron-bombarded chemically etched W tips.

**Theoretical calculation**

First-principles calculations were based on the generalized gradient approximation[9] using the full-potential projector-augmented wave method[10], as implemented in the Vienna *ab initio* simulation package (VASP)[11]. Experimental structural parameters were used in the calculations[12]. Electronic structures of bulk $SrVO_3$ were calculated using a $20 \times 20 \times 20$ Monkhorst–Pack k-mesh over the Brillouin zone (BZ). To simulate surface electronic structures, a $\sqrt{2} \times \sqrt{2} \times 11$ supercell that included apical oxygen was used. Crystal relaxation was permitted during the calculations. A tight-binding model Hamilton was constructed by projecting onto the Wannier orbitals[13-15], which used the VASP2WANNIER90 interface[16]. The V atom's *d* orbitals and O atom's *p* orbitals were used to construct Wannier functions without maximizing localization.

Combining the bulk and surface Wannier functions, the surface spectral weight of a semi-infinite (001) slab was calculated using an iterative Green's function method. The bulk and surface Wannier functions were obtained from the inner layer and the surface of the slab, respectively. The surface spectral weights $A(k)$ were obtained using the surface Green's functions. The scattering probability was obtained using the formula $I(q) = \int A(k) \times A(k+q) dk$ to simulate the quasiparticle interference (QPI) pattern. Note that the surface spectral weight obtained from the iterative Green's function method includes contributions from the surface-projected bulk states as well as the surface states, so that both these contributions are included in our QPI simulations.

**Negligible sample dependence of QPI patterns**

Three samples with thicknesses of $t$ = 110 nm, 40 nm, and 20 nm, were prepared. The *c*-axis lattice constants of the three samples were identical and comparable to that of bulk $SrVO_3$[1]. This suggests that the films were relaxed, which was also supported by reciprocal lattice mapping (Fig. S2f). All three films had ($\sqrt{2}\times\sqrt{2}$)-*R*45° reconstructed surfaces (Fig. 1). Within this thickness range, we could not observe thickness-dependent QPI patterns (Figs. 3e, f and Fig. S5).

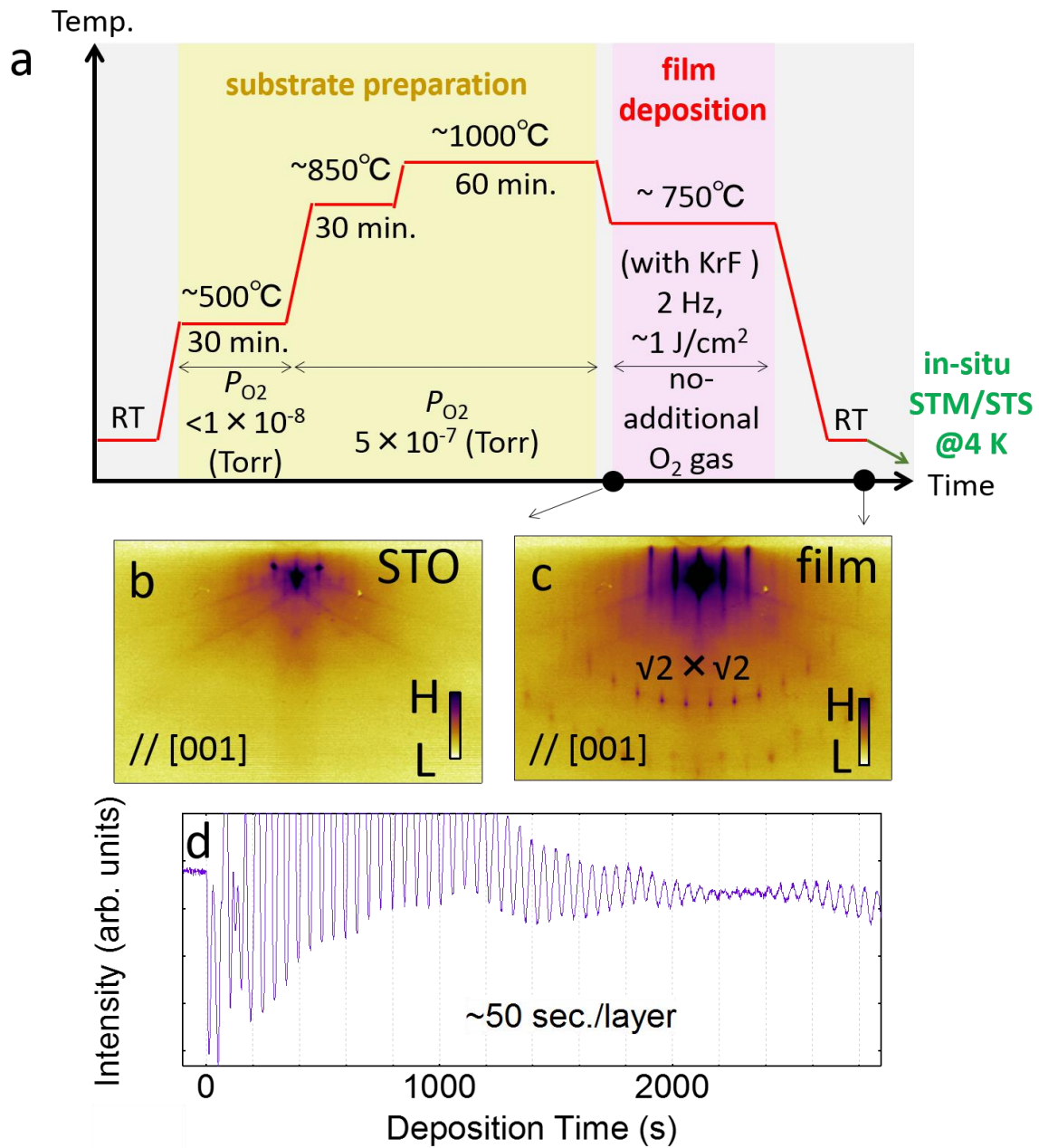

**Figure S1 | Thin film growth of SrVO$_3$(001) on Nb-doped SrTiO$_3$(001). a,** Temperature profile acquired during the deposition process. **b, c,** Typical reflection high-energy electron diffraction (RHEED) patterns from the Nb-doped SrTiO$_3$(001) substrate before film deposition (**b**) and from the film after deposition (**c**) along the [001] direction. **d,** Typical RHEED intensity oscillations during the initial stage of the film deposition process.

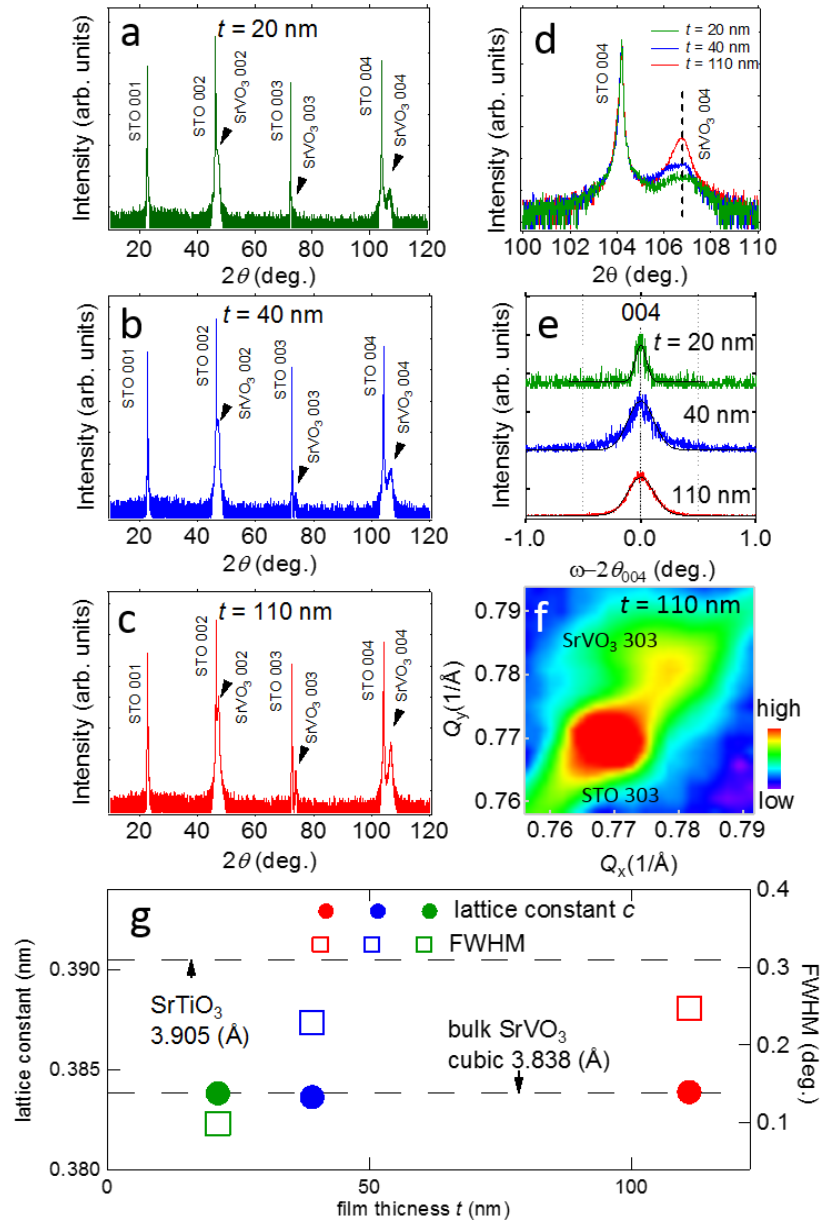

**Figure S2 | *Ex situ* XRD characterization. a–c**, Out-of-plane XRD patterns for samples with film thicknesses of $t = 20$ nm (green), 40 nm (blue), and 110 nm (red) on SrTiO$_3$ substrates. **d, e**, XRD patterns near the 004 peak of the substrate and films (**d**) and rocking curves of the 004 peaks (**e**). Full-width-at-half-maximum (FWHM) values were determined by fitting the data with a Gaussian. **f,** Reciprocal-space mapping near the 303 reflection of the substrate with a 110-nm-thick film. **g,** The thicknesses *t*, lattice constants *c,* and rocking-curve FWHM values for three samples.

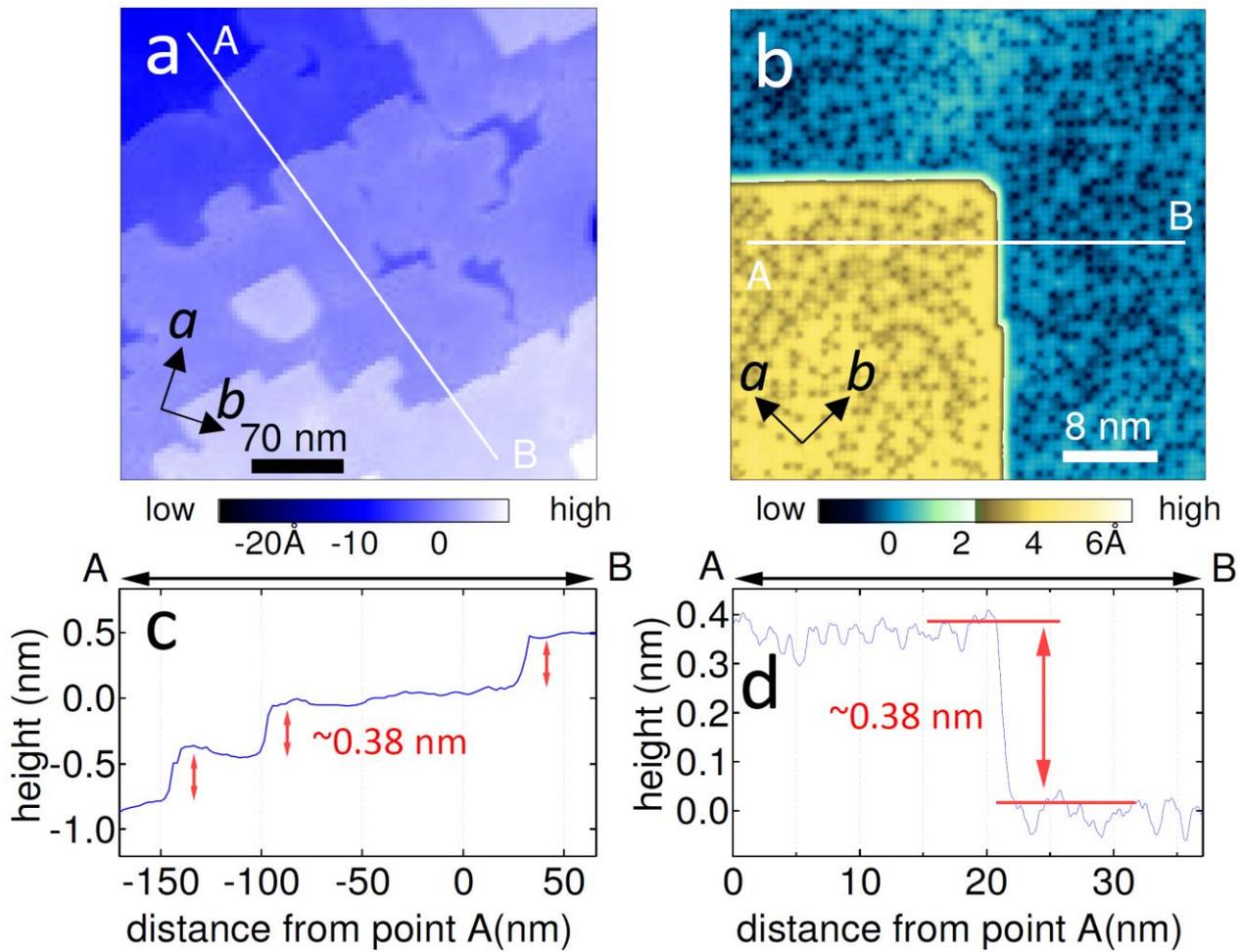

**Figure S3 | Typical STM images of SrVO$_3$(001) film surfaces. a, b**, A wide-area topographic image (**a**) and a high-resolution topographic image across a single step-edge (**b**) (sample bias voltage $V_s = -100$ meV). **c, d**, Height evolutions along the line A–B in **a** and **b** are shown in **c** and **d**, respectively. A typical height difference across the steps was approximately 0.38 nm (red arrows in **c** and **d**), which corresponds to the *c*-axis lattice constant of SrVO$_3$.

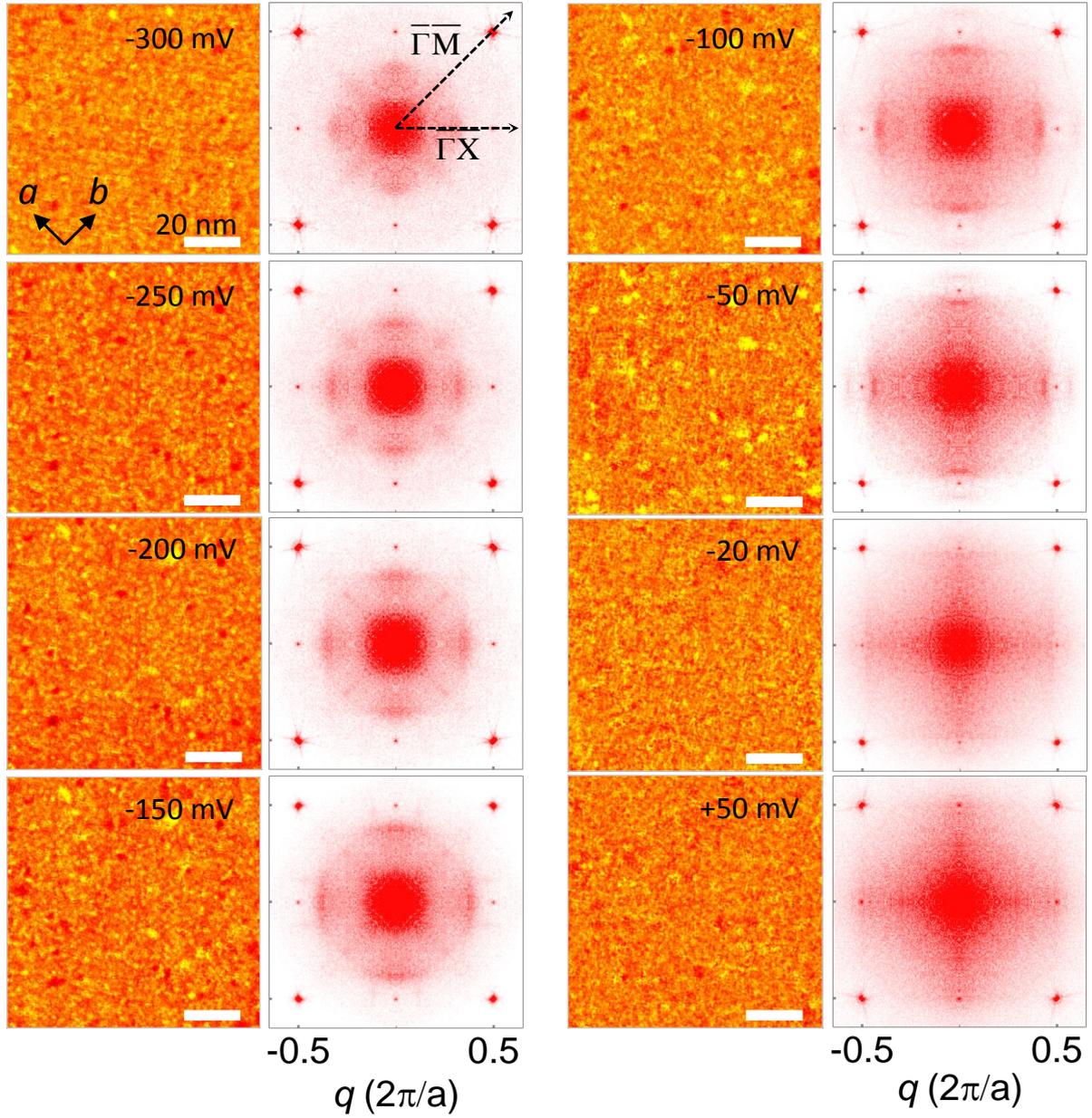

**Figure S4 | Energy evolution of conductance maps and their FTs (110-nm-thick film).** Conductions maps obtained from the same location as in **Fig. 2a**. FT images are rotated and four-fold symmetrized. Note that standing momentum *q\** is along the direction of the nearest-neighbor V atoms (along *a*- and *b*-direction), which we define as the $\overline{\Gamma}\overline{X}$ direction (see Fig. 1c).

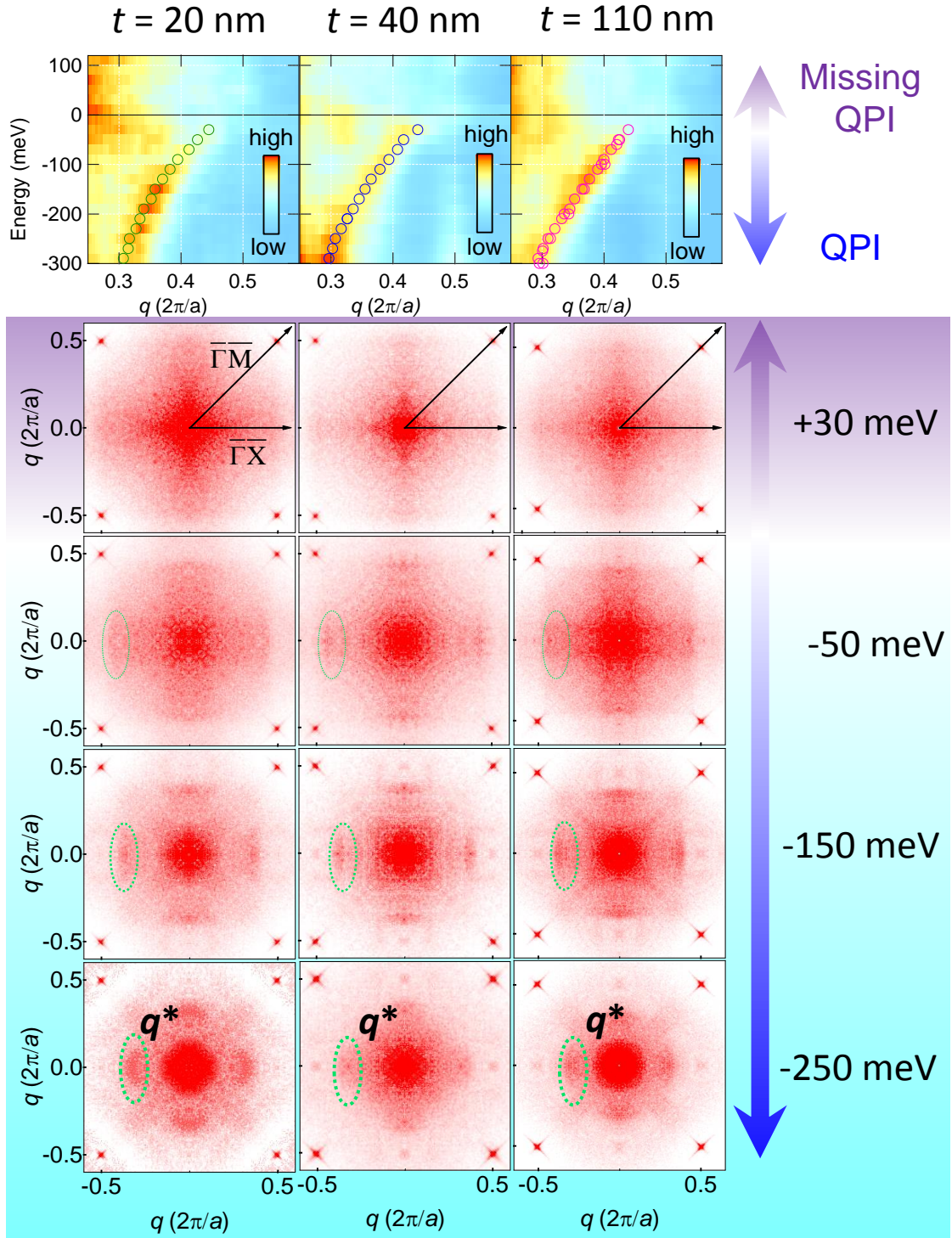

**Figure S5 | Energy evolution of the QPI patterns for samples with thicknesses of 20 nm, 40 nm, and 110 nm.** Energy evolution for the QPI is similar for the three samples. FT images are rotated and four-fold symmetrized. Note that standing momentum $q^*$ is along the direction of the nearest-neighbor V atoms (along *a*- and *b*-direction), which we define as the $\overline{\Gamma X}$ direction (see Fig. 1c).

| Thickness (*t*) | Lattice constant (*c*) | FWHM of the 004 peak | Corresponding Figures (STM data) |
|---|---|---|---|
| 20 nm | 0.3838 nm | 0.099° | Figs 3 e, f<br>Fig. S2<br>Fig. S5 (right) |
| 40 nm | 0.3836 nm | 0.229° | Figs 3 e,f<br>Fig. S1e<br>Fig. S2<br>Fig. S5 (mid) |
| 110 nm | 0.3839 nm | 0.247° | Figs 1 a, b, d<br>Figs 2 a, b<br>Figs 3 a–d<br>Figs 3 e, f<br>Figs S1b–d<br>Fig. S2<br>Fig. S3<br>Fig. S4<br>Fig. S5 (left) |

**Table S1 | Experimental sample properties.**

| Orbital | Bulk energy | $V_O$-site energy (*with* apical O atom) | $V_b$-site energy (*without* apical O atom) |
|---|---|---|---|
| $d_{xz}, d_{yz}$ | 0.1253 eV | 1.3025 eV | 0.2639 eV |
| $d_{xy}$ | 0.1253 eV | 0.2981 eV | 0.4597 eV |
| $d_{z^2}$ | 2.7583 eV | 3.2064 eV | 1.5886 eV |
| $d_{x^2-y^2}$ | 2.7583 eV | 2.2645 eV | 3.2391 eV |

**Table S2 | Onsite energies for the five orbitals studied for the surface and bulk.**